\documentclass[a4paper,12pt]{article}

\usepackage[english]{babel}
\usepackage[a4paper,tmargin=3truecm,bmargin=3truecm,rmargin=2.5truecm,
lmargin=2.5truecm,twoside,verbose=true]{geometry}

\usepackage{cancel,graphicx}
\usepackage[dvips]{hyperref}

\usepackage{amsmath,amssymb}

\numberwithin{equation}{section}

\allowdisplaybreaks[1]

\renewenvironment{thebibliography}[1]
         {\section*{References}\frenchspacing\small
          \begin{list}{[\arabic{enumi}]}
         {\usecounter{enumi}\parsep=2pt\topsep 0pt
         \settowidth{\labelwidth}{[#1]}
         \leftmargin=\labelwidth\advance\leftmargin\labelsep
         \rightmargin=0pt\itemsep=1pt\sloppy}}{\end{list}}

\title{On the Effective Action of Noncommutative Yang-Mills Theory\footnote{At the occasion of the ``International Conference on Noncommutative Geometry and Physics'', April 2007, Orsay (France). Work supported by ANR grant NT05-3-43374 ``GenoPhy''.}}
\author{Axel de Goursac}
\date{}

\begin{document}

\maketitle
\vspace*{-1cm}
\begin{center}
\textit{Laboratoire de Physique Th\'eorique, B\^at.\ 210\\
    Universit\'e Paris XI,  F-91405 Orsay Cedex, France }\\[1ex]
\textit{Mathematisches Institut der Westf\"alischen
  Wilhelms-Universit\"at \\Einsteinstra\ss{}e 62, D-48149 M\"unster,
  Germany\\
  e-mail: \texttt{axelmg@melix.net} }\\
\end{center}%

\vskip 2cm

\begin{abstract}
We compute here the Yang-Mills effective action on Moyal space by integrating over the scalar fields in a noncommutative scalar field theory with harmonic term, minimally coupled to an external gauge potential. We also explain the special regularisation scheme chosen here and give some links to the Schwinger parametric representation. Finally, we discuss the results obtained: a noncommutative possibly renormalisable Yang-Mills theory.
\end{abstract}%

\pagebreak

\section{Introduction.}

\subsection{Motivations.}

Recently, there has been an intense activity on the noncommutative field theories (see \cite{Douglas:2001ba,Szabo}). Noncommutative geometry \cite{CONNES,CM} is indeed one of the most attractive candidates for new physics beyond the Standard Model. Within the framework of noncommutative geometry, it is possible to deform usual ``commutative spaces'' in ``noncommutative spaces''. They are in fact related to noncommutative associative algebras, interpreted as algebras of functions on these ``noncommutative spaces''. The Moyal space, as a deformation of $\mathbb R^D$, is one of the simplest, but field theories can be constructed on other ``noncommutative spaces'' (see for example \cite{Gayral:2006wu}). However, field theories defined on Moyal space \cite{Wulkenhaar:2006si,vince} can be seen as a limiting regime of string theory \cite{Seiberg:1999vs,Witten:1985cc} and matrix theory \cite{Connes:1997cr}. Noncommutative geometry can also be used to describe physics in the presence of a background field thanks to non-local interactions, like the quantum Hall effect \cite{hall1,hall2,hall3}.

The simplest generalisation of scalar $\phi^4$ field theory to Moyal space was known to suffer from a new type of divergence, the UV/IR mixing \cite{Minwalla:1999px,Chepelev:1999tt}, which is responsible for the non-renormalisability of this model. To cure this problem, Grosse and Wulkenhaar proposed a few years ago to add some harmonic potential in the action \cite{Grosse:2004yu,Grosse:2003aj}. The renormalisability of this new theory has been proven at any order of perturbation in the matrix base \cite{Grosse:2004yu,Grosse:2003nw}, by multi-scale analysis \cite{Rivasseau:2005gd}, in the $x$-space \cite{Gurau:2005gd} and using dimensional regularisation \cite{Tanasa:2007qx}. Furthermore, this noncommutative field theory seems to have a better flow behavior than the commutative $\phi^4$ model. Indeed, contrary to the commutative model, the Grosse-Wulkenhaar theory does not present a Landau ghost \cite{beta1,beta23,beta}; and a constructive version is possible \cite{constructiva1,constructiva2}. Then, other renormalisable noncommutative field theories have been obtained, for example the LSZ theory \cite{Langmann:2003if,Langmann:2003cg}, which is a scalar theory covariant under the Langmann-Szabo duality \cite{Langmann:2002cc} like the Grosse-Wulkenhaar model, or the scalar $\phi^3$ model \cite{Grosse:2005ig,Grosse:2006qv,Grosse:2006tc}. The noncommutative version of the fermionic Gross-Neveu model is also renormalisable \cite{Vignes-Tourneret:2006nb,Vignes-Tourneret:2006xa,Lakhoua:2007ra}, even if it still suffers from the UV/IR mixing.

As in the case of noncommutative scalar field theory, the naive generalisation of Yang-Mills theory to Moyal space, whose action is $\int F_{\mu\nu}\star F_{\mu\nu}$, exhibits the UV/IR mixing divergence \cite{Matusis:2000jf}, making its renormalisability quite problematic. To ensure the renormalisability of Yang-Mills theory, one has to find what could be the analog of the harmonic term for a gauge theory. One possible way is to compute the Yang-Mills effective action from a noncommutative scalar field theory with harmonic term coupled with an external gauge field \cite{de Goursac:2007gq,Grosse:2007qx}. The present paper is describing such a way.

The paper is organised as follows. In the next subsection, we present the Moyal space, the noncommutative scalar field theory with harmonic term, and the gauge theory in the framework of noncommutative geometry. Then, in section 2, the minimal coupling between the scalar field theory and an external gauge potential is given. We recall the definition of the effective action and the computation \cite{de Goursac:2007gq}. The connexions between this calculation and the parametric representation \cite{Rivasseau:2006qx} is further presented in subsection 2.3. In subsection 2.4, we explain the special regularisation scheme used in the computation. Finally, the effective action gives rise to a gauge theory, which is a possible candidate for renormalisability. This is discussed in section 3.

\subsection{Framework.}

We recall here the definition of the Moyal space, the noncommutative scalar $\phi^4_4$ field theory with harmonic term, and the gauge theory over the Moyal space \cite{Gracia-Bondia:1987kw,Varilly:1988jk}. We do not give many mathematical details \cite{de Goursac:2007gq,Wallet:2007em}. The Moyal space is the deformation of Euclidean $\mathbb{R}^4$ associated to the Moyal-Groenenwald product
\begin{align}
\forall f,h\in\mathcal{S}\quad(f\star h)(x)=\frac{1}{\pi^4\theta^4}\int d^4y\,d^4z\ 
f(x+y)\,h(x+z)e^{-iy\wedge z},\label{eq:moyal}
\end{align}
where $\mathcal S=\mathcal S(\mathbb R^4)$ is the space of complex-valued Schwartz functions on $\mathbb R^4$, $x\wedge y=2x_\mu\Theta^{-1}_{\mu\nu}y_\nu$ and
\begin{align}
 \Theta_{\mu\nu}=\theta\begin{pmatrix} 0 & -1 & 0 & 0 \\ 1 & 0 & 0 & 0 \\ 0 & 0 & 0 & -1 \\ 0 & 0 & 1 & 0 \end{pmatrix} . \label{eq:theta}
 \end{align}
Notice that $\forall f,h\in\mathcal S$, $f\star h\in\mathcal S$. In fact, we can extend this product to certain distributions upon using duality of linear spaces, and we define the Moyal algebra as a subspace of tempered distributions $\mathcal S'(\mathbb R^4)$:
\begin{align}
\mathcal M=\{T\in\mathcal S'(\mathbb R^4),\quad \forall f\in\mathcal S\quad T\star f\in\mathcal S\quad \text{and}\quad f\star T\in\mathcal S\}.
\end{align}

The Moyal algebra involves in particular the ``coordinate functions'' $x_\mu$, satisfying $[x_\mu,x_\nu]_\star =x_\mu\star x_\nu-x_\nu\star x_\mu=i\Theta_{\mu\nu}$. There are some important properties of $\mathcal M$:
\begin{subequations}
\begin{align}
\forall f,h\in\mathcal M,\qquad &(f\star h)^\dag=h^\dag\star f^\dag,\qquad 
\partial_\mu(f\star h)=\partial_\mu f\star h+f\star\partial_\mu h, \label{eq:relat1}\\
&\int d^4x\ f\star h=\int d^4x\ f.h,\label{eq:relat2}\\
&[\widetilde x_\mu,f]_\star=2i\partial_\mu f,\qquad \{\widetilde x_\mu,f\}_\star=\widetilde x_\mu\star\widetilde x_\nu+\widetilde x_\nu\star\widetilde x_\mu=2\widetilde x_\mu.f ,\label{eq:relat3}
\end{align}
\end{subequations}
where $\widetilde x_\mu=2\Theta^{-1}_{\mu\nu}x_\nu$.

The noncommutative complex scalar orientable $\phi^4_4$ field theory with harmonic term is constructed on this Moyal space and its action is given by
\begin{align}
S(\phi)=\int d^4x\big(\partial_\mu\phi^\dag\star\partial_\mu\phi
+\Omega^2(\widetilde{x}_\mu\phi)^\dag\star(\widetilde{x}_\mu\phi)
+m^2\phi^\dag\star\phi+\lambda \phi^\dag\star\phi\star\phi^\dag\star\phi\big),\label{eq:actionharm}
\end{align}
where $\phi$ is a complex scalar field of mass $m$, and the parameters $\Omega$ and $\lambda$ are dimensionless. This theory is renormalisable for any value of $\Omega$. The constraint $\Omega\neq0$ is needed for the renormalisability proof only in the case of non-orientable interactions ($\phi^\dag\star\phi^\dag\star\phi\star\phi$) of complex-valued fields \cite{Grosse:2004yu,Gurau:2005gd}.

We give here the expression of the propagator $C(x,y)$ \cite{Gurau:2005qm} and of the interaction vertex
\begin{subequations}
\begin{align}
C(x,y)=\frac{\Omega^2}{\pi^2\theta^2}\int_0^\infty \!\! 
\frac{dt}{\sinh^2(2{\widetilde{\Omega}}t)} & \exp\Big(
-\frac{{\widetilde{\Omega}}}{4}\coth({\widetilde{\Omega}}t)(x{-}y)^2
-\frac{{\widetilde{\Omega}}}{4}\tanh({\widetilde{\Omega}}t)(x{+}y)^2-m^2t\Big),
\label{eq:propag} \\
\int d^4x(\phi^\dag\star\phi\star\phi^\dag\star\phi)(x) &=\frac{1}{\pi^4\theta^4}\int\prod_{i=1}^4d^4x_i\,\phi^\dag(x_1)\phi(x_2)\phi^\dag(x_3)\phi(x_4)\nonumber\\
&\times\delta(x_1-x_2+x_3-x_4)e^{-i\sum_{i<j}(-1)^{i+j+1}x_i\wedge x_j } \label{eq:interaction},
\end{align}
\end{subequations}
where we have defined ${\widetilde{\Omega}}= 2\frac{\Omega}{\theta}$.

In the Yang-Mills theory, fields are gauge potentials, associated to connections. We can therefore define the notion of (noncommutative) connection \cite{CONNES,Dubois-Violette:1989vq,de Goursac:2007gq,Wallet:2007em} on the Moyal space:
\begin{align}
 \forall\phi\in\mathcal M,\quad \nabla_\mu\phi=\partial_\mu\phi-iA_\mu\star\phi, \label{eq:nabla}
\end{align} 
where $A_\mu$ is a real field in $\mathcal M$, the gauge potential associated to the connection $\nabla_\mu$. Then, the group of gauge transformations acts on the different fields as
\begin{subequations}
\begin{align}
\phi^g &=g\star\phi\label{eq:gaugephi},\\
A_\mu^g &=g\star A_\mu\star g^\dag+ig\star\partial_\mu g^\dag\label{eq:gaugepot},
\end{align}
\end{subequations}
where $g\in\mathcal M$ is the gauge function and it satisfies $g^\dag\star g=g\star g^\dag=\mathbb I$.

An important feature of the Moyal space is that it involves an invariant connection. It turns out that 
\begin{align}
 \xi_\mu=-\frac{1}{2}\widetilde{x}_\mu \label{eq:invarcon} 
\end{align}
defines a connection which is invariant under gauge transformations. The occurrence of such invariant connections is not new in noncommutative geometry and has been already mentionned in earlier studies focused in particular on matrix-valued field theories \cite{Dubois-Violette:1989vq,Dubois-Violette:1998,Masson:1999,Masson:2005}. Indeed, because of the equation \eqref{eq:relat3}, which can be reexpressed in $\partial_\mu\phi=[i\xi_\mu,\phi]_\star$, we can compute that
\begin{align}
\xi_\mu^g=g\star \xi_\mu\star g^\dag+ig\star\partial_\mu g^\dag=\xi_\mu.\label{eq:invar-xi}
\end{align}
From this special connection, we construct the ``covariant coordinates'' \cite{Douglas:2001ba}:
\begin{align}
\mathcal A_\mu=A_\mu-\xi_\mu.\label{eq:covcoord}
\end{align}
This field $\mathcal A_\mu$ is not a gauge potential but a covariant field: 
\begin{align}
\mathcal A^g_\mu=g\star\mathcal A_\mu\star g^\dag.\label{eq:gaugecov}
\end{align}
Furthermore, it is possible to express the curvature $F_{\mu\nu}=\partial_\mu A_\nu-\partial_\nu A_\mu-i [A_\mu,A_\nu]_\star$, which is also covariant, $F^g_{\mu\nu}=g\star F_{\mu\nu}\star g^\dag$, in terms of $\mathcal A_\mu$:
\begin{align}
F_{\mu\nu}=\Theta^{-1}_{\mu\nu}-i[\mathcal A_\mu,\mathcal A_\nu]_\star.\label{eq:Fcovcoord}
\end{align}

\section{The one-loop order effective action.}

\subsection{Definition.}

The whole formalism needed in this article is presented in subsection 1.2. We can now start to define and later to compute the effective action for the Yang-Mills theory. We couple in a first time the above action \eqref{eq:actionharm} with the gauge potential $A_\mu$ in order to get a gauge-invariant action. Notice that this coupled action does not contain any kinetic term in $A_\mu$, just couplings with the scalar field $\phi$. Then we can integrate over the quantized fields $\phi$, considering $A_\mu$ as an external field. We compute this integration at the one-loop order and get what is called the effective action in $A_\mu$. This effective action contains now some kinetic terms in $A_\mu$, but no more term involving $\phi$. This effective action is in fact a candidate for a Yang-Mills theory, and this candidate is interesting because if we couple this Yang-Mills theory to a scalar field whose action is \eqref{eq:actionharm}, this coupled action will be stable at the quantum level. This property of stability is important for the proof of the renormalisability of a field theory.

Let us now define a ``minimal coupling'' between the scalar field $\phi$ and the gauge potential $A_\mu$. With the following prescriptions
\begin{subequations}
\begin{align}
\partial_\mu\phi&\mapsto \partial_\mu\phi-iA_\mu\star\phi, \label{eq:coup1}\\
\widetilde x_\mu\phi &\mapsto \widetilde x_\mu\phi+A_\mu\star\phi,\label{eq:coup2}
\end{align}
\end{subequations}
we replace in the scalar action \eqref{eq:actionharm} the gauge non-invariant terms $\partial_\mu\phi$ and $\widetilde x_\mu\phi$ by gauge covariant terms, and we get the coupled action
\begin{align}
S(\phi,A)=\int d^4x\  \big(&(\partial_\mu\phi-iA_\mu\star\phi)^\dag\star(\partial_\mu\phi-iA_\mu\star\phi)
+\Omega^2(\widetilde{x}_\mu\phi+A_\mu\star\phi)^\dag\star(\widetilde{x}_\mu\phi+A_\mu\star\phi)\nonumber\\
& +m^2\phi^\dag\star\phi+\lambda \phi^\dag\star\phi\star\phi^\dag\star\phi\big),\label{eq:actioncoupled}
\end{align}
which is gauge invariant. Using equations \eqref{eq:relat3}, we can simplify this action:
\begin{align}
S(\phi,A) = S(\phi)+ \int d^4x \  \big(&(1+\Omega^2)\phi^\dag\star
(\widetilde{x}_\mu A_\mu)\star\phi-(1-\Omega^2)\phi^\dag\star A_\mu \star\phi\star
\widetilde{x}_\mu\nonumber\\
& +(1+\Omega^2)\phi^\dag\star A_\mu\star
A_\mu\star \phi\big).\label{eq:harmcoupled}
\end{align}

\begin{figure}[!htb]
  \centering
  \includegraphics[scale=1]{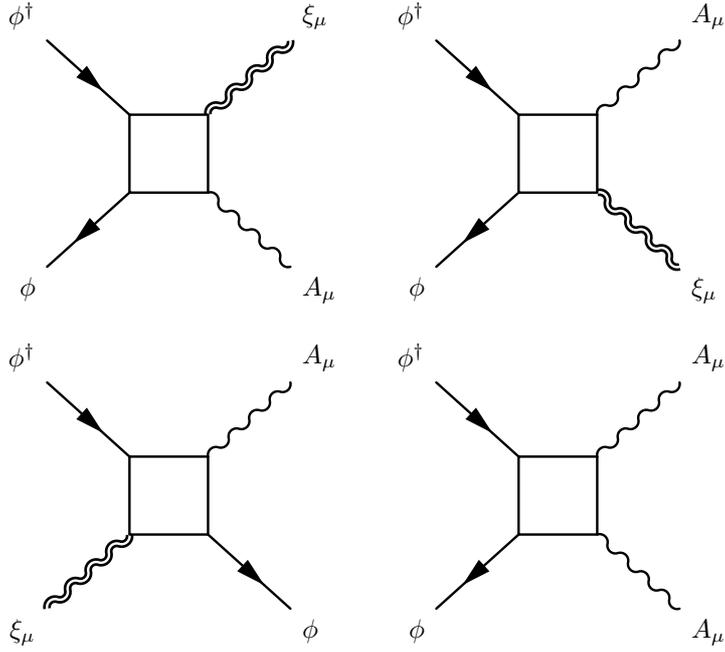}
  \caption[The vertices]{\footnotesize{Graphical representation for
      the vertices carrying the external gauge potential $A_\mu$
      involved in the action \eqref{eq:harmcoupled}. The overall
      factor affecting the two uppermost vertices is $(1 + \Omega^2)$.
      From left to right, the overall factors affecting the lower
      vertices are respectively equal to $- 2(1 - \Omega^2)$ and $- (1
      + \Omega^2)$.}}
  \label{fig:vertices}
\end{figure}

From the coupled action \eqref{eq:harmcoupled}, we define the effective action $\Gamma(A)$.
\begin{align}
e^{-\Gamma(A)}= \int D\phi D\phi^\dag e^{-S(\phi,A)}. \label{eq:defact}
\end{align}
Notice that the interaction part $\int \phi^\dag\star\phi\star\phi^\dag\star\phi$ of $S(\phi)$ \eqref{eq:actionharm} does not occur in the computation of the effective action at the one-loop order. Nevertheless, there are some additional vertices involving $A_\mu$ and/or $\xi_\mu$ (or equivalently $\widetilde x_\mu$) and
generated by the minimal coupling, which can be obtained by combining
\eqref{eq:harmcoupled} with the generic
relation
\begin{align}
\int d^4x(f_1\star f_2\star f_3\star f_4)(x)=\frac{1}{\pi^4\theta^4}\int 
\prod_{i=1}^4d^4x_i\, f_1(x_1) f_2(x_2) f_3(x_3) f_4(x_4) \nonumber\\
\times\delta(x_1-x_2+x_3-x_4)e^{-i\sum_{i<j}(-1)^{i+j+1}x_i\wedge x_j }.\label{eq:exprvertex}
\end{align}
These vertices are depicted on the Figure \ref{fig:vertices}. Note
that additional overall factors must be taken into account. These are
indicated on the Figure \ref{fig:vertices}.

\subsection{Computation.}

\begin{figure}[!htb]
  \centering
  \includegraphics[scale=1]{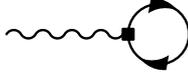}
  \caption[1-point]{\footnotesize{The non vanishing tadpole diagram.
      To simplify the figure, we do not explicitly draw all the
      diagrams that would be obtained from the vertices given on the
      figure 2 but indicate only the overall topology of the
      corresponding diagrams. Notice that the background lines are not
      explicitly depicted.}}
  \label{fig:1-point}
\end{figure}

\begin{figure}[!htb]
  \centering
  \includegraphics[scale=1]{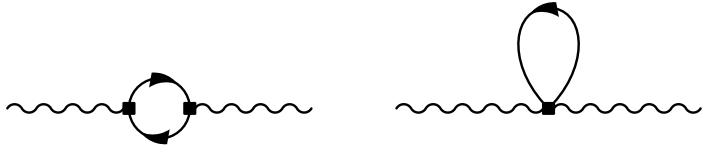}
  \caption[Two-point]{\footnotesize{Relevant one-loop diagrams
      contributing to the two-point function. To simplify the figure,
      we do not explicitly draw all the diagrams that would be
      obtained from the vertices given in figure 2 but indicate
      only the overall topology of the corresponding diagrams. Notice
      that the background lines are not explicitly depicted. The
      leftmost (resp.\ rightmost) diagram corresponds to the
      contribution $\mathcal{T}_2'$ (resp.\ $\mathcal{T}_2''$).}}
  \label{fig:2-point}
\end{figure}

\begin{figure}[!htb]
  \centering
  \includegraphics[scale=1]{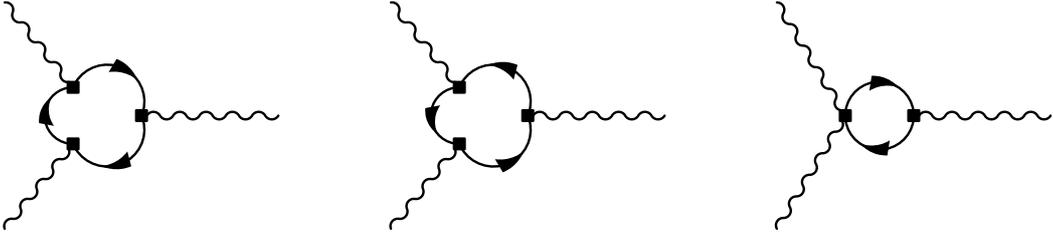}
  \caption[3-point]{\footnotesize{Relevant one-loop diagrams
      contributing to the 3-point function. Comments similar to those
      related to the figure 4 apply. The rightmost (resp.\ two
      leftmost) diagram(s) corresponds to the contribution
      $\mathcal{T}_3''$ (resp.\ $\mathcal{T}_3'$).}}
  \label{fig:3-point}
\end{figure}

\begin{figure}[!htb]
  \centering
  \includegraphics[scale=1]{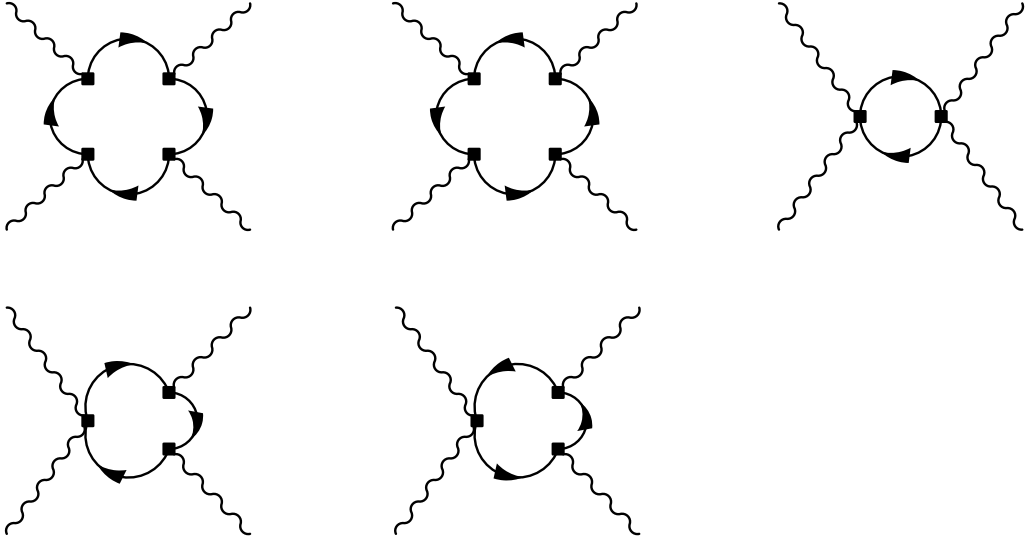}
  \caption[4-point]{\footnotesize{Relevant one-loop diagrams
      contributing to the 4-point function. Comments similar to those
      related to the figure 4 apply. Among the upper figures, the
      rightmost figure (resp.\ the two leftmost) diagram(s)
      corresponds to the contribution $\mathcal{T}_4'''$ (resp.\
      $\mathcal{T}_4'$). The lower diagrams correspond to
      $\mathcal{T}_4''$. }}
  \label{fig:4-point}
\end{figure}

We now compute all the contributions of the effective action $\Gamma(A)$. The corresponding diagrams of these contributions are depicted on the Figures \ref{fig:1-point}-\ref{fig:4-point}.

The expression for the vertices \eqref{eq:exprvertex} permits one to obtain the amplitude corresponding to the tadpole on the Figure \ref{fig:1-point}
\begin{align}
\mathcal{T}_1= \frac{1}{\pi^4\theta^4}\int d^4x\ d^4u\ d^4z\ A_\mu(x)\  
e^{-i(x-u)\wedge z}\ C(u+z,u)\  
((1-\Omega^2)(2\widetilde{u}_\mu+\widetilde{z}_\mu)-2\widetilde{x}_\mu). 
\label{eq:tadpole1}
\end{align}
Using the expression of the propagator \eqref{eq:propag}, \eqref{eq:tadpole1} becomes
\begin{align}
\mathcal{T}_1=& \frac{\Omega^2}{4\pi^6\theta^6}\int d^4x\ d^4u\ d^4z 
\int_0^\infty  \frac{dt\ e^{-tm^2}}{ \sinh^2(\widetilde{\Omega}t)
\cosh^2(\widetilde{\Omega}t)}\  A_\mu(x)\  e^{-i(x-u)\wedge z}\nonumber\\
&\times e^{-\frac{\widetilde{\Omega}}{4} (\coth(\widetilde{\Omega}t)z^2
+\tanh(\widetilde{\Omega}t)(2u+z)^2}
((1-\Omega^2)(2\widetilde{u}_\mu+ \widetilde{z}_\mu)-2\widetilde{x}_\mu).
\label{eq:tadpole2}
\end{align}
To simplify the calculations, we can introduce the following 8-dimensional vectors $X$, $J$ and the $8\times 8$ matrix $K$ defined by
\begin{align}
X=\begin{pmatrix} u\\ z \end{pmatrix}, \quad 
K=\begin{pmatrix} 4\tanh(\widetilde{\Omega}t) \mathbb{I} & 
2\tanh(\widetilde{\Omega}t)\mathbb{I} -2i\Theta^{-1} \\  
2\tanh(\widetilde{\Omega}t)\mathbb{I} +2i\Theta^{-1} &  
(\tanh(\widetilde{\Omega}t)+ \coth(\widetilde{\Omega}t))\mathbb{I}
\end{pmatrix} ,\quad
\ J=\begin{pmatrix} 0\\ i\widetilde{x} \end{pmatrix}. \label{eq:tadpole2bis}
\end{align}
Then, using these new variables, equation \eqref{eq:tadpole2} can be conveniently reexpressed in a form such that some Gaussian integrals can be easily performed. Notice that this procedure can be adapted to the calculation of the higher order Green functions. The combination of \eqref{eq:tadpole2bis} with \eqref{eq:tadpole2} yields
\begin{align}
\mathcal{T}_1=& \frac{\Omega^2}{4\pi^6\theta^6}
\int d^4x\ d^4u\ d^4z \int_0^\infty  
\frac{dt\ e^{-tm^2}}{
  \sinh^2(\widetilde{\Omega}t)\cosh^2(\widetilde{\Omega}t)}\  
A_\mu(x) \nonumber\\
&\times e^{-\frac{1}{2}X.K.X+J.X} 
((1-\Omega^2)(2\widetilde{u}_\mu+ \widetilde{z}_\mu)-2\widetilde{x}_\mu).
\end{align}
Evaluating the Gaussian integral over $X$, we find
\begin{align}
\mathcal{T}_1=-\frac{\Omega^4}{\pi^2\theta^2(1+\Omega^2)^3}
\int d^4x \int_0^\infty  \frac{dt\ e^{-tm^2}}{ 
\sinh^2(\widetilde{\Omega}t)\cosh^2(\widetilde{\Omega}t)}\ 
A_\mu(x)\widetilde{x}_\mu\ e^{-\frac{2\Omega}{\theta(1+\Omega^2)}
\tanh(\widetilde{\Omega}t)x^2}. \label{eq:tadpole3}
\end{align}
This latter expression has a quadratic and a logarithmic UV divergence for $t \to 0$. To obtain explicitly these divergences, we make a Taylor expansion of \eqref{eq:tadpole3}
\begin{align}
\mathcal{T}_1 =& \int d^4x\ \Big(-\frac{\Omega^2}{4\pi^2(1+\Omega^2)^3\epsilon}
 \widetilde{x}_\mu A_\mu -\frac{m^2\Omega^2\ln(\epsilon)}{4\pi^2(1+\Omega^2)^3}\widetilde{x}_\mu A_\mu
 -\frac{\Omega^4\ln(\epsilon)}{\pi^2\theta^2(1+\Omega^2)^4} x^2\widetilde{x}_\mu A_\mu \Big)+\dots,\label{eq:tadpole4}
\end{align}
where $\epsilon \to 0$ is a cut-off and the ellipses denote finite
contributions. We can see that the tadpole \eqref{eq:tadpole4} is non-vanishing. This is a rather unusual feature for a Yang-Mills theory, which will be discussed in section 3.

To compute the 2, 3 and 4-point functions, we can apply the same procedure to the one used for the tadpole: matrix and vector change of variables, Gaussian integration and Taylor expansion for the variables $t_i\to0$ of the different propagators which occur in these amplitudes. We just give here the final expressions for the various contributions.
The contributions corresponding to the diagrams of the Figure \ref{fig:2-point} can be expressed as
\begin{subequations}
\begin{align}
\mathcal{T}_2' &= \int \! d^4x\Big(\frac{(1{-}\Omega^2)^2}{16\pi^2(1{+}\Omega^2)^3\epsilon}
A_\mu A_\mu + \frac{m^2(1{-}\Omega^2)^2\ln(\epsilon)}{16\pi^2(1{+}\Omega^2)^3} A_\mu A_\mu +\frac{\Omega^2(1{-}\Omega^2)^2\ln(\epsilon)}{4\pi^2\theta^2(1{+}\Omega^2)^4} x^2A_\mu(x)A_\mu(x)\nonumber\\
&- \frac{\Omega^4\ln(\epsilon)}{2\pi^2(1{+}\Omega^2)^4}(\widetilde{x}_\mu A_\mu)^2  -\frac{(1{-}\Omega^2)^2(1{+}4\Omega^2{+}\Omega^4)\ln(\epsilon)}{96\pi^2(1{+}\Omega^2)^4}A_\mu\partial^2 A_\mu -\frac{(1{-}\Omega^2)^4\ln(\epsilon)}{96\pi^2(1{+}\Omega^2)^4}(\partial_\mu A_\mu)^2\Big)+ \dots,\label{eq:2pt}\\
\mathcal{T}_2'' &= \int \!d^4x\Big(-\frac{1}{16\pi^2(1{+}\Omega^2)\epsilon} A_\mu A_\mu
- \frac{m^2\ln(\epsilon)}{16\pi^2(1{+}\Omega^2)} A_\mu A_\mu -\frac{\Omega^2\ln(\epsilon)}{4\pi^2\theta^2(1{+}\Omega^2)^2} x^2A_\mu A_\mu\nonumber\\ &+\frac{\Omega^2\ln(\epsilon)}{16\pi^2(1{+}\Omega^2)^2} A_\mu\partial^2 A_\mu\Big)  + \dots
\end{align}
\end{subequations}
Note that for $\mathcal T_2'$, we used a special scheme of regularisation, $\int_{\epsilon/4}^\infty$ instead of $\int_{\epsilon}^\infty$, which will be discussed in subsection 2.4.

The computation of the 3-point function contributions can be
conveniently carried out by further using the following identity
\begin{align}
\int \!d^4x\ \widetilde{x}_\mu A_\mu (A_\nu\star A_\nu)=\frac{1}{2}\int \! d^4x\ 
\Big(\widetilde{x}_\mu A_\nu\{A_\mu,A_\nu\}_\star \; 
-i(\partial_\mu A_\nu)[A_\mu,A_\nu]_\star\Big).\label{eq:3pid}
\end{align}
With the expression of \eqref{eq:3pid}, the 3-point contributions depicted on the Figure \ref{fig:3-point} are given by
\begin{subequations}
\begin{align}
\mathcal{T}_3' &= \int \! d^4x\Big(
\frac{\Omega^2(1-\Omega^2)^2\ln(\epsilon)}{8\pi^2(1+\Omega^2)^4}\widetilde{x}_\mu A_\nu\{A_\mu,A_\nu\}_\star\nonumber\\
&-\frac{i(1-\Omega^2)^2(1+4\Omega^2+\Omega^4)\ln(\epsilon)}{
48\pi^2(1+\Omega^2)^4}(\partial_\mu A_\nu)[A_\mu,A_\nu]_\star\Big)+ \dots, \\
\mathcal{T}_3'' &= \int \!d^4x\Big( -\frac{\Omega^2\ln(\epsilon)}{8\pi^2(1{+}\Omega^2)^2}
\widetilde{x}_\mu A_\nu\{A_\mu,A_\nu\}_\star +\frac{i\Omega^2\ln(\epsilon)}{8\pi^2(1{+}\Omega^2)^2}
(\partial_\mu A_\nu)[A_\mu,A_\nu]_\star\Big)+ \dots
\end{align}
\end{subequations}

In the same way, the final expressions for the diagrams on the Figure \ref{fig:4-point} are
\begin{subequations}
\begin{align}
\mathcal{T}_4' &= -\frac{(1{-}\Omega^2)^4\ln(\epsilon)}{96\pi^2(1{+}\Omega^2)^4} 
\int \!d^4x\ \Big( (A_\mu\star A_\nu)^2
+2(A_\mu\star A_\mu)^2 \Big)+ \dots, \\
\mathcal{T}_4'' &= \frac{(1{-}\Omega^2)^2\ln(\epsilon)}{16\pi^2(1{+}\Omega^2)^2}
\int \!d^4x\ (A_\mu\star A_\mu)^2+ \dots, \\
\mathcal{T}_4''' &= -\frac{\ln(\epsilon)}{32\pi^2}\int \! d^4x\ 
(A_\mu\star A_\mu)^2 + \dots\label{eq:t4}
\end{align}
\end{subequations}

Finally, we collect the various contributions given above, using the following identities
\begin{subequations}
\begin{align}
\int \!d^4x\ {\mathcal{A}}_\mu\star{\mathcal{A}}_\mu 
&=\int \!d^4x\Big(\frac{1}{4}{\widetilde{x}}^2+{\widetilde{x}}_\mu A_\mu
+A_\mu A_\mu\Big),
\\
\int \!d^4x\ F_{\mu\nu}\star F_{\mu\nu}
&= \int d^4x\Big(-2(A_\mu\partial^2A_\mu+(\partial_\mu A_\mu)^2)
-4i\partial_\mu A_\nu[A_\mu,A_\nu]_\star-[A_\mu,A_\nu]_\star^2 \Big),
\\
\int \!d^4x\; \{{\mathcal{A}}_\mu,{\mathcal{A}}_\nu\}^2_\star
&=\int \!d^4x\; \Big(\frac{1}{4}({\widetilde{x}}^2)^2
+2{\widetilde{x}}^2{\widetilde{x}}_\mu A_\mu
+4({\widetilde{x}}_\mu A_\mu)^2+2{\widetilde{x}}^2A_\mu A_\mu
\nonumber
\\
&+2(\partial_\mu A_\mu)^2
+4{\widetilde{x}}_\mu A_\nu\{A_\mu,A_\nu\}_\star
+\{A_\mu,A_\nu\}^2_\star\Big),
\end{align}
\end{subequations}
and we find the expression for the effective action $\Gamma(A)$
\begin{align}
\Gamma(A) &= \int \!d^4x\Big[\frac{\Omega^2}{4\pi^2(1{+}\Omega^2)^3}\left(\frac{1}{\epsilon}+m^2\ln(\epsilon)\right)
\left(\mathcal{A}_\mu\star\mathcal{A}_\mu -\frac{1}{4}\widetilde{x}^2\right) -\frac{(1{-}\Omega^2)^4\ln(\epsilon)}{192\pi^2(1{+}\Omega^2)^4} F_{\mu\nu}\star F_{\mu\nu}\nonumber\\
 & +\frac{\Omega^4\ln(\epsilon)}{8\pi^2(1{+}\Omega^2)^4}\left(F_{\mu\nu}\star  F_{\mu\nu}
+\{\mathcal{A}_\mu, \mathcal{A}_\nu\}_\star^2 
-\frac{1}{4}(\widetilde{x}^2)^2\right)\Big]+ \dots ,
 \label{eq:zegamma}
\end{align}
where $\mathcal{A}_\mu(x)= A_\mu(x)+\frac 1 2\widetilde{x}_\mu$ and $F_{\mu\nu}=\partial_\mu A_\nu-\partial_\nu A_\mu-i[A_\mu,A_\nu]_\star$.

\subsection{Relation with the parametric representation.}

We will see in this subsection how the above calculations could have been done in the parametric representation of the noncommutative scalar field theory (see \cite{Rivasseau:2006qx,Rivasseau:2007qx,Tanasa:2007qx,param,mellin}). Let us recall in a first time some basic features of this parametric representation of the four-dimensional theory defined by the action \eqref{eq:actionharm}. The propagator \eqref{eq:propag} can be reformulated in ($\widetilde \Omega=\frac{2\Omega}{\theta}$)
\begin{align}
C(x,y)&=\frac{\Omega}{4\pi^2\theta}\int_0^\infty \!\! \frac{d\alpha}{\sinh^2(\alpha)} e^{-\frac{m^2\alpha}{2\widetilde \Omega}}C(x,y,\alpha),\\
C(x,y,\alpha) &= \exp\Big(-\frac{\widetilde \Omega}{4}\coth(\frac \alpha 2)(x-y)^2 -\frac{\widetilde \Omega}{4}\tanh(\frac \alpha 2)(x+y)^2\Big).
\end{align}
Through the following identity
\begin{align}
\delta(x_1-x_2+x_3-x_4)=\int\frac{d^4p}{\pi^4\theta^4}e^{-ip\wedge(x_1-x_2+x_3-x_4)},
\end{align}
the vertex \eqref{eq:interaction} becomes
\begin{align}
\int d^4x(\phi^\dag\star\phi\star\phi^\dag\star\phi)(x) &=\frac{1}{\pi^8\theta^8}\int\prod_{i=1}^4d^4x_i\,d^4p\,\phi^\dag(x_1)\phi(x_2)\phi^\dag(x_3)\phi(x_4) V(x_1,x_2,x_3,x_4,p),\\
V(x_1,x_2,x_3,x_4,p)&=\exp\left(-i\sum_{i<j}(-1)^{i+j+1}x_i\wedge x_j-i\sum_{i}(-1)^{i+1}p\wedge x_i\right).
\end{align}

Consider now a graph $G$ with a set $V$ of $n$ internal vertices, $N$ external legs and a set $L$ of $(2n-\frac N 2)$ internal lines or propagators. In the theory defined by \eqref{eq:actionharm}, there are four positions (called ``corners'') associated to each vertex $v\in V$, and each corner is bearing either a half internal line or an external field. These corners are denoted by $x^v_i$, where $i\in\{1,..,4\}$ is given by the cyclic order of the Moyal product. The set $I\subset V\times\{1,..,4\}$ of internal corners (hooked to some internal line) has $4n-N$ elements whereas the set $E=V\times\{1,..,4\}\setminus I$ of external corners has $N$ elements. Each vertex $v\in V$ carries also a hypermomentum, which is noted $p_v$. A line $l\in L$ of the graph $G$ joins two corners in $I$, and we note their positions by $x^{l,1}$ and $x^{l,2}$. We will also note the external corners $x_e$. Be careful that each corner has two notations for its position. In these notations, we can express the amplitude $\mathcal A_G$ of such a graph $G$
\begin{align}
\mathcal A_G(\{x_e\})=&\left(\frac{\Omega}{4\pi^2\theta}\right)^{2n-\frac N 2}\left(\frac{1}{\pi^8\theta^8}\right)^n \int_0^\infty \prod_{l\in L}\frac{d\alpha_l}{\sinh^2(\alpha_l)} \int \prod_{(v,i)\in I}d^4x^v_i\prod_{v\in V}d^4p_v \nonumber\\
&\times\prod_{l\in L}C(x^{l,1},x^{l,2},\alpha_l)\prod_{v\in V} V(x^v_1,x^v_2,x^v_3,x^v_4,p_v).
\end{align}
We took here $m^2=0$ to simplify calculations. The article \cite{Rivasseau:2006qx} dealing with the parametric representation tells us that by performing Gaussian integrations, we get a Gaussian function of the external variables, divided by a determinant. If we note $t_l=\tanh(\frac{\alpha_l}{2})$, using $\sinh(\alpha_l)=\frac{2t_l}{1-t_l^2}$, then the expression of the amplitude is given by
\begin{align}
\mathcal A_G(\{x_e\})=&\left(\frac{\theta}{8\Omega}\right)^{2n-\frac N 2}\left(\frac{1}{\pi^4\theta^4}\right)^n \int_0^\infty \prod_{l\in L}(d\alpha_l(1-t_l^2)^2) HU_G(t)^{-2}e^{-\frac{HV_G(x_e,t)}{HU_G(t)}},\label{eq:polydef}
\end{align}
and that stands for a definition of the polynome $HU_G(t)$ in the $t$ variables and the quadratic form $HV_G(x_e,t)$ in the external variables $x_e$, which is also polynomial in the $t$ variables. Let us now recall the computation of $HU$ and $HV$.
\begin{align}
\mathcal A_G(\{x_e\})=&\left(\frac{\Omega}{16\pi^2\theta}\right)^{2n-\frac N 2}\left(\frac{1}{\pi^8\theta^8}\right)^n \int_0^\infty \prod_{l\in L}\frac{d\alpha_l(1-t_l^2)^2}{t_l^2} \int \prod_{(v,i)\in I}d^4x^v_i\prod_{v\in V}d^4p_v \nonumber\\
&\times\prod_{l\in L} e^{-\frac{\widetilde \Omega}{4t_l}(x^{l,1}-x^{l,2})^2 -\frac{\widetilde \Omega t_l}{4}(x^{l,1}+x^{l,2})^2} \prod_{v\in V} V(x^v_1,x^v_2,x^v_3,x^v_4,p_v).
\end{align}

At this point, we introduce the short variables $u$ and the long ones $v$
\begin{align}
u_l=&\frac{x^{l,1}-x^{l,2}}{\sqrt 2}\nonumber\\
v_l=&\frac{x^{l,1}+x^{l,2}}{\sqrt 2}
\end{align}
associated to each line $l\in L$. Notice that the Jacobian of the transformation is 1. Each internal line $l\in L$ of the graph $G$ joins two vertices (or one two times). This fact will be expressed by the $((2n-\frac N 2)\times 4)$-dimensional incidence matrix $\epsilon^v$ associated to each vertex $v\in V$. We define $\epsilon^v_{li}=(-1)^{i+1}$ if the line $l\in L$ hooks the vertex $v$ at corner $i\in\{1,..,4\}$, $\epsilon^v_{li}=0$ if not, and $\eta^v_{li}=|\epsilon^v_{li}|$. Then the short and long variables are given by
\begin{align}
u_l=\frac{1}{\sqrt 2}\sum_{v\in V}\sum_{i=1}^4\epsilon^v_{li}x^v_i\nonumber\\
v_l=\frac{1}{\sqrt 2}\sum_{v\in V}\sum_{i=1}^4\eta^v_{li}x^v_i,
\end{align}
and for $(v,i)\in I$ (and not for $(v,i)\in E$),
\begin{align}
x^v_i=\frac{1}{\sqrt 2}\sum_{l\in L}(\epsilon^v_{li}u_l+\eta^v_{li}v_l).
\end{align}
The vector $\chi_i^v=1$ if $(v,i)\in E$ and $\chi_i^v=0$ if not, shows how the external legs are hooked on vertices. We define also $\omega_{ij}=1$ if $i<j$, $\omega_{ij}=-1$ if $j>i$ and $\omega_{ii}=0$. The amplitude is then given by
\begin{align}
\mathcal A_G(\{x_e\})=&\left(\frac{\Omega}{16\pi^2\theta}\right)^{2n-\frac N 2}\left(\frac{1}{\pi^8\theta^8}\right)^n \int_0^\infty \prod_{l\in L}\frac{d\alpha_l(1-t_l^2)^2}{t_l^2} \int \prod_{l\in L}d^4u_ld^4v_l\prod_{v\in V}d^4p_v \nonumber\\
&\prod_{l\in L} e^{-\frac{\widetilde \Omega}{2t_l}u_l^2 -\frac{\widetilde \Omega t_l}{2}v_l^2} \prod_{v\in V} \exp\Big(-i\sum_{i<j}\frac{(-1)^{i+j+1}}{2}\sum_{l,l'\in L} (\epsilon^v_{li}u_l+\eta^v_{li}v_l)\wedge (\epsilon^v_{l'j}u_{l'}+\eta^v_{l'j}v_{l'})\nonumber\\
& -i\sum_{i}\frac{(-1)^{i+1}}{\sqrt 2}\sum_{l\in L} p_v\wedge (\epsilon^v_{li}u_l+\eta^v_{li}v_l) -i\sum_{i\neq j}\frac{(-1)^{i+j+1}}{\sqrt 2}\sum_{l\in L} \chi^v_i\omega_{ij}x^v_i\wedge (\epsilon^v_{lj}u_l+\eta^v_{lj}v_l)\nonumber\\
& -i\sum_{i}(-1)^{i+1}\chi^v_i p_v\wedge x^v_i -i\sum_{i<j}(-1)^{i+j+1} \chi^v_i\chi_j^v x^v_i\wedge x^v_j\Big).\label{eq:bigexp}
\end{align}
The sums over $i$ and $j$ are between 1 and 4. We note
\begin{align}
X=\begin{pmatrix} x_e \\ u \\ v \\ p \end{pmatrix}\quad\text{and}\quad G=\begin{pmatrix} M & P \\ P^T & Q \end{pmatrix},
\end{align}
where $M$ is a $4N\times 4N$ matrix (quadratic form for external variables), $Q$ is a $[8(2n-\frac N 2)+4n]\times [8(2n-\frac N 2)+4n]$ matrix (quadratic form for short and long variables and hypermomenta), and $P$ is the coupling. The following expression together with \eqref{eq:bigexp} define $M$, $P$ and $Q$,
\begin{align}
\mathcal A_G(\{x_e\})=&\left(\frac{\Omega}{16\pi^2\theta}\right)^{2n-\frac N 2}\left(\frac{1}{\pi^8\theta^8}\right)^n \int_0^\infty \frac{d\alpha(1-t^2)^2}{t^2} \int dudvdp\, e^{-\frac 1 2 X^TGX}.
\end{align}
By performing Gaussian integrations, we obtain
\begin{align}
\mathcal A_G(\{x_e\})=&\left(\frac{\pi^2\Omega}{\theta}\right)^{2n-\frac N 2}\left(\frac{4}{\pi^6\theta^8}\right)^n \int_0^\infty d\alpha(1-t^2)^2\, \frac{e^{-\frac 1 2 (x_e)^T(M-PQ^{-1}P^T)(x_e)}}{\sqrt{\det Q}}.
\end{align}
We deduce from the latter equation the expression of polynoms $HU_G(t)$ and $HV_G(x_e,t)$ defined in \eqref{eq:polydef}.

\begin{figure}[!htb]
  \centering
  \includegraphics[scale=1]{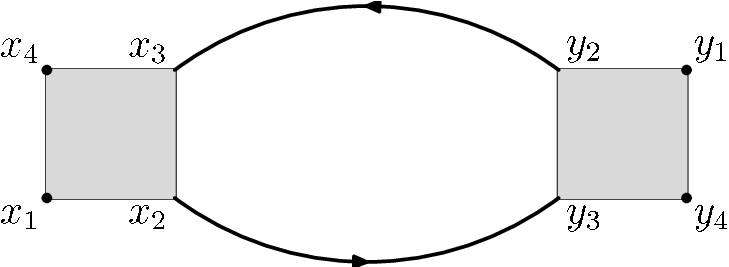}
  \caption[bubble]{\footnotesize{The bubble graph.}}
  \label{fig:bubble}
\end{figure}

Let us now apply this formalism to the calculation of the Yang-Mills effective action. It would be possible to compute all the contributions to the effective action $\Gamma(A)$ we found in subsection 2.2. We will nevertheless reduce our study to the contribution $\mathcal T_4'''$. One can notice on the Figures \ref{fig:4-point} and \ref{fig:vertices} that this contribution involves only the ``bubble graph'' (see Figure \ref{fig:bubble}).
In \cite{Rivasseau:2006qx}, it was found the expression for the polynom $HU_G(t)$ and the real part of $HV_G(x_e,t)$ of the ``bubble graph''. Notice that we have integrated over the hypermomentum of the rooted vertex (see \cite{Rivasseau:2006qx}). We have computed here the imaginary part of $HV_G(x_e,t)$ and we have obtained
\begin{align}
HU_G(t)=&4(t_1+t_2)^2,\\
HV_G(x_e,t)=&\frac{2(1+\Omega^2)}{\Omega\theta}(t_1+t_2)((x_1-x_4+y_1-y_4)^2+t_1t_2(x_1-x_4-y_1+y_4)^2)\nonumber\\
&-4i(t_1^2-t_2^2)(x_1-x_4)\wedge (y_1-y_4)-4i(t_1+t_2)^2(x_1\wedge x_4+y_1\wedge y_4).
\end{align}
As a consequence, the amplitude $\mathcal T_4'''$ is given by
\begin{align}
\mathcal T_4'''=& K \int_{\epsilon'}^\infty \frac{d\alpha_1 d\alpha_2(1-t_1^2)^2(1-t_2^2)^2}{16(t_1+t_2)^4} \int d^4x_1d^4x_4d^4y_1d^4y_4 A_\mu(x_1)A_\mu(x_4)A_\nu(y_1)A_\nu(y_4)\nonumber\\ &\exp\Big(-\frac{m^2\theta(\alpha_1+\alpha_2)}{4\Omega}-\frac{1+\Omega^2}{2\Omega\theta(t_1+t_2)}((x_1-x_4+y_1-y_4)^2
+t_1t_2(x_1-x_4-y_1+y_4)^2)\nonumber\\
&+i\frac{t_1-t_2}{t_1+t_2}(x_1-x_4)\wedge (y_1-y_4)+i(x_1\wedge x_4+y_1\wedge y_4) \Big),
\end{align}
where $K=\frac{2(1+\Omega^2)^2}{64\Omega^2\pi^8\theta^6}$ and $\epsilon'=2\widetilde \Omega \epsilon$ an UV cut-off. In the expression of $K$, $2(1+\Omega^2)^2$ comes from the combinatory of $\mathcal T_4'''$ and the overall factors affecting the vertices (see Figure \ref{fig:vertices}), whereas the other part of $K$ comes from the overall factor of the equation \eqref{eq:polydef}. If we keep only the leading term in $\epsilon'$, we get
\begin{align}
\mathcal T_4'''= K \int_{\epsilon'}^\infty & \frac{d\alpha_1 d\alpha_2}{16(t_1+t_2)^4}\int d^4x_1d^4x_4d^4y_1d^4y_4 A_\mu(x_1)A_\mu(x_4)A_\nu(y_1)A_\nu(y_4)\nonumber\\
&\exp\Big(-\frac{1+\Omega^2}{2\Omega\theta(t_1+t_2)}(x_1-x_4+y_1-y_4)^2
+i\frac{t_1-t_2}{t_1+t_2}(x_1-x_4)\wedge (y_1-y_4)\nonumber\\
&+i(x_1\wedge x_4+y_1\wedge y_4) \Big).
\end{align}
Now we can make a change of variables $y_4=x_1-x_4+y_1+z$, and a Taylor expansion of $A_\nu(y_4)$
\begin{align}
\mathcal T_4'''=& K \int_{\epsilon'}^\infty \frac{d\alpha_1 d\alpha_2}{16(t_1+t_2)^4}\int d^4x_1d^4x_4d^4y_1d^4z A_\mu(x_1)A_\mu(x_4)A_\nu(y_1)(A_\nu(x_1-x_4+y_1)+O_{z\to0}(z)_\nu)\nonumber\\
&\exp\Big(-\frac{1+\Omega^2}{2\Omega\theta(t_1+t_2)}z^2
-i\frac{t_1-t_2}{t_1+t_2}(x_1-x_4)\wedge z+i(x_1\wedge x_4+y_1\wedge x_1-y_1\wedge x_4+y_1\wedge z) \Big).\label{eq:reg1}
\end{align}
By performing a Gaussian integration over $z$ together with a Taylor expansion on the variable $\epsilon'\to0$ at the first order, the amplitude becomes
\begin{align}
\mathcal T_4'''= K' \int_{\epsilon'}^\infty &\frac{d\alpha_1 d\alpha_2}{(t_1+t_2)^2}\int d^4x_1d^4x_4d^4y_1 A_\mu(x_1)A_\mu(x_4)A_\nu(y_1)\nonumber\\
&(A_\nu(x_1-x_4+y_1)+O_{(t_1,t_2)\to(0,0)}(t_1,t_2)_\nu)e^{+i(x_1\wedge x_4+y_1\wedge x_1-y_1\wedge x_4)},
\end{align}
where $K'=\frac{\Omega^2\pi^2\theta^2}{4(1+\Omega^2)^2}K$. If we simplify this expression, we obtain
\begin{align}
\mathcal T_4'''= -\frac{\ln(\epsilon)}{32\pi^2} \int d^4x (A_\mu\star A_\mu\star A_\nu\star A_\nu)(x)+O_{\epsilon\to0}(1),
\end{align}
and we recognize the equation \eqref{eq:t4}. This was just an example to show the calculation power of the parametric representation in our Yang-Mills effective action's computation.

\subsection{Gauge invariance and regularisation scheme.}

In this subsection, we shall explain and justify the special regularisation scheme used in the calculations (see $\mathcal T_2'$ in subsection 2.2). In \cite{Gayral:2004cs}, the one-loop effective action can be expressed in terms of heat kernels
\begin{align}
\Gamma_{1l}[\phi,A]=\frac 12 \ln(\det(H(\phi,A)H(0,0)^{-1})),\label{eq:defact2}
\end{align}
where $H(\phi,A)=\frac{\delta^2 S(\phi,A)}{\delta \phi \,\delta \phi^\dag}$ is the effective potential. In the Schwinger representation, one obtains
\begin{align}
\Gamma_{1l}[\phi,A]&=-\frac{1}{2} \int_0^\infty \frac{dt}{t} \,
\mathrm{Tr}\big( e^{-t  H(\phi,A)} - e^{-t H(0,0)}\big)\\
&= -\frac{1}{2} \lim_{s \to 0} \Gamma(s)\,\mathrm{Tr}\big(
H^{-s}(\phi,A)-H^{-s}(0,0)\big).
\end{align}
By using $\Gamma(s+1)=s\Gamma(s)$ and expanding \cite{Connes:2006qj}
\begin{align}
H^{-s}(\phi,A) =\big(1+a_1(\phi,A) s + a_2(\phi,A) s^2 + \dots\big)  
H^{-s}(0,0),
\end{align}
one obtains the following expression for the effective action
\begin{align}
\Gamma_{1l}[\phi,A]
= -\frac{1}{2} \lim_{s \to 0} \mathrm{Tr}\Big(\big(
\Gamma(s+1)a_1(\phi,A) + s \Gamma(s+1) a_2(\phi,A) 
+\dots \big) H^{-s}(0,0)\Big).\label{eq:hkact}
\end{align}
The expansion $\Gamma(s+1)=1-s\gamma+\dots$ permits one to reexpress \eqref{eq:hkact} in the form
\begin{align}
\Gamma_{1l}[\phi,A]
&= -\frac{1}{2} \lim_{s \to 0} \mathrm{Tr}\big(
a_1(\phi,A) H^{-s}(0,0)\big) 
\nonumber
\\
&-\frac{1}{2} \lim_{s \to 0} \mathrm{Tr} 
\Big( s\big(a_2(\phi,A) -\gamma a_1(\phi,A)\big)  H^{-s}(0,0)\Big).
\end{align}
We can reconcile the both definitions of the effective action \eqref{eq:defact} and \eqref{eq:defact2} by $\Gamma(A)=\Gamma_{1l}[0,A]$. So, we get
\begin{align}
\Gamma(A)&= -\frac{1}{2} \lim_{s \to 0} \mathrm{Tr}\big(a_1(A) H^{-s}(0,0)\big) 
\nonumber
\\
&-\frac{1}{2} \lim_{s \to 0} \mathrm{Tr} 
\Big( s\big(a_2(A) -\gamma a_1(A)\big)  H^{-s}(0,0)\Big),
\end{align}
where the ellipses denote finite contributions. The second part of the effective action is called the Wodzicki residue \cite{Wodzicki:1984} and corresponds to the logarithmically divergent part of $\Gamma(A)$. This residue is a trace and is gauge invariant. But the first part of $\Gamma(A)$, which corresponds to the quadratically divergent part, is not gauge invariant. That is why the naive $\epsilon$-regularisation of the Schwinger integrals breaks the gauge invariance of the theory in the quadratically divergent part. One possible way to solve this problem could be the restoration of gauge invariance by using methods from algebraic renormalisation \cite{Piguet:1995er}. Notice also that a dimensional regularisation scheme could have been used \cite{Tanasa:2007qx}, where the divergent contributions of the effective action were automatically gauge invariant. A Hopf algebra description of this type of renormalisation, the Tanasa-Vignes algebra, has been given in \cite{hopfnc} for the noncommutative scalar field theories.

We give here another way, which works for this special type of theory and which has been used for the calculations (see subsection 2.2). To find a convenient regularisation scheme, one can do the following transformation on the cut-off $\epsilon\to\lambda\epsilon$, where $\lambda\in\mathbb R_+^\ast$, and it can be different for each contribution. The logarithmically divergent part of the effective action is insensitive to a finite scaling of the cut-off, so these transformations will affect only the quadratically divergent part, which is not gauge invariant.

Let us consider the contribution of a graph $G_p$ with $p$ internal lines and suppose that it involves a quadratic divergence. If not, this contribution does not need any change for the restoration of the gauge invariance of the effective action. Using the polynom $HU_{G_p}$ \cite{Rivasseau:2006qx} and integrating on the appropriated variables like in \eqref{eq:reg1}, one obtains at the one-loop order that the quadratically divergent part is proportional (after Taylor expansion on $\epsilon\to 0$) to $\int_\epsilon^1\frac{dt_i}{(t_1+...+t_p)^{p+1}}$. Indeed the integration on the $t_i$ between 1 and $+\infty$ gives only finite contributions. The expression of the propagator \eqref{eq:propag} implies that a part of the logarithmic divergence is proportional to $-m^2\int_\epsilon^1\frac{dt_i}{(t_1+...+t_p)^{p}}$ with exactly the same factor as the quadratic divergence. As
\begin{align}
\int_\epsilon^1\frac{dt_i}{(t_1+...+t_p)^{p+1}}&=\frac{1}{p(p!)\epsilon}+\dots,\nonumber\\
\int_\epsilon^1\frac{dt_i}{(t_1+...+t_p)^{p}}&=-\frac{\ln(\epsilon)}{(p-1)!}+\dots,
\end{align}
the contribution of $G_p$ writes $\frac{1}{p(p!)\epsilon}K_p+\frac{m^2\ln(\epsilon)}{(p-1)!}K_p+\dots$, where the ellipses denote other logarithmic contributions and finite terms, and $K_p$ is a factor depending on the graph $G_p$. The logarithmically divergent term involving $m^2$ in the effective action is gauge invariant. That is why one has to choose a regularisation scheme $\epsilon\to\lambda_p\epsilon$ so that $\frac{1}{p(p!)\lambda_p}=\frac{1}{(p-1)!}$, i.e. $\lambda_p=\frac{1}{p^2}$.

Therefore, with the regularisation scheme $\epsilon\to\frac{\epsilon}{p^2}$ for a graph with $p$ internal lines, one can conclude that the gauge invariance is restored in the effective action. Indeed, we made the transformation $\epsilon\to\frac{\epsilon}{4}$ for the $\mathcal T_2'$ contribution (see \eqref{eq:2pt}). Finally, notice that this method is very specific to this theory and to the one-loop order. As noted above, the algebraic renormalisation or the dimensional regularisation provide more general results.

\section{Discussion.}

Let us now summarise the above results. We coupled a scalar theory with gauge potentials to obtain a gauge invariant action in subsection 2.1. We then integrated this action over scalar fields and got the one-loop effective action of Yang-Mills theory. We showed the procedure to compute this effective action in subsection 2.2 and noticed the links with the Schwinger parametric representation in subsection 2.3. Within this parametric representation, the calculations are indeed easier. We discussed also the special scheme of regularisation used in the computations, and explained the reasons why we chose it in subsection 2.4.

At this point, we obtain from \eqref{eq:zegamma} a gauge invariant action \cite{de Goursac:2007gq}
\begin{align}
S=\int d^4x \Big(\frac{1}{4g^2}F_{\mu\nu}\star F_{\mu\nu}
+\frac{\Omega'^2}{4g^2}\{\mathcal{A}_\mu,\mathcal{A}_\nu\}^2_\star
+\frac{\kappa}{2} {\mathcal{A}}_\mu\star{\mathcal{A}}_\mu
\Big)\label{eq:result}
\end{align}
in term of the gauge potential $A_\mu$, which is stable at the quantum level if we couple this Yang-Mills theory to some scalar theory with harmonic term. The scalar theory of Grosse-Wulkenhaar is renormalisable to all orders of perturbation, because the harmonic term solves the problem of UV/IR mixing of this scalar theory \cite{Grosse:2004yu,Gurau:2005gd}. As the pure Yang-Mills theory, given by $S=\int F_{\mu\nu}\star F_{\mu\nu}$, suffers from the same problem of UV/IR mixing \cite{Matusis:2000jf}, it is tempting to conjecture that \eqref{eq:result} is a good candidate for renormalisability. Notice that another group arrived at similar conclusions by a different method \cite{Grosse:2007qx}, and that the action \eqref{eq:result} can be derived from a spectral action principle \cite{Chamseddine:1996zu,Grosse:2007jy}. Indeed, taking $B_\mu=A_\mu$ and $\phi=0$, the action of \cite{Grosse:2007jy} can be reexpressed in the form of \eqref{eq:result}.

Let us go to the details of this action \eqref{eq:result}. Beyond the usual $\int F_{\mu\nu}\star F_{\mu\nu}$ Yang-Mills contribution, we find additional gauge invariant terms of quadratic and quartic order in $\mathcal A_\mu$ (which can be reexpressed in term of $A_\mu$ through \eqref{eq:covcoord}) $\int \mathcal A_\mu\star\mathcal A_\mu$ and $\int\{\mathcal{A}_\mu,\mathcal{A}_\nu\}^2_\star$. The quadratic term involves a mass term for the field $A_\mu$, while such a bare mass term is forbidden by gauge invariance in Yang-Mills theories on commutative spaces. In this noncommutative theory, it is therefore possible to have some massive gauge fields without breaking the gauge invariance of the theory. The quartic term $\int\{\mathcal{A}_\mu,\mathcal{A}_\nu\}^2_\star$ may be viewed as the gauge counterpart of the harmonic term $\int\widetilde x^2\varphi^2=\int\{\widetilde x_\mu,\varphi\}^2_\star$, solving the problem of UV/IR mixing and ensuring the renormalisability of the $\varphi^4$ theory investigated in \cite{Grosse:2004yu}.

But before the study of this quantum field theory and its renormalisability, there is already an unusual feature at the level of the classical field theory. We noticed in the subsection 2.2 that the tadpole was not vanishing. This is the sign of a non-vanishing vacuum expectation value for the field $A_\mu$, which is the direct consequence of the presence of the quadratic and quartic term in $\mathcal A_\mu$ in \eqref{eq:result}. Let us have a look on the equation of the motion
\begin{align}
\frac{1}{g^2}[\mathcal{A}_\nu,[\mathcal{A}_\nu,\mathcal{A}_\mu]_\star]_\star +\frac{\Omega'^2}{g^2}\{\mathcal{A}_\nu,\{\mathcal{A}_\nu,\mathcal{A}_\mu\}_\star\}_\star +2\kappa\mathcal{A}_\mu=0. \label{eq:motion}
\end{align}
Notice that this equation can be reexpressed in term of $A_\mu=\mathcal A_\mu-\frac 1 2\widetilde x_\mu$. We see indeed that $A_\mu=0$ (which is equivalent to $\mathcal A_\mu=\frac 1 2\widetilde x_\mu$) is not a solution of the equation \eqref{eq:motion} for the real parameters $\Omega'^2\neq0$ or $\kappa\neq0$. Nonetheless, $A_\mu=-\frac 1 2\widetilde x_\mu$ (which is equivalent to $\mathcal A_\mu=0$) is solution of the equation of the motion \eqref{eq:motion} for all $\Omega'^2$ and $\kappa$. By expanding the action \eqref{eq:result} around this solution $\mathcal A_\mu=0$ through \eqref{eq:Fcovcoord}, we find the following action (up to an inessential constant term)
\begin{align}
S=\int d^4x \Big(\frac \kappa 2 \mathcal A_\mu\star\mathcal A_\mu+ \frac{(1+\Omega'^2)}{2g^2}\mathcal A_\mu\star\mathcal A_\mu\star\mathcal A_\nu\star\mathcal A_\nu -\frac{(1-\Omega'^2)}{2g^2}\mathcal A_\mu\star\mathcal A_\nu\star\mathcal A_\mu\star\mathcal A_\nu\Big),\label{eq:actionexp}
\end{align}
where $\mathcal A_\mu$ is then the quantum gauge field with vanishing expectation value. Note that for $\kappa>0$, the propagator $\frac 2 \kappa \delta_{\mu\nu}\delta(x-y)$, which is trivial, is non-negative, so $\mathcal A_\mu=0$ is a minimum of the action \eqref{eq:result}. But then, as the action \eqref{eq:actionexp} involves only this trivial quadratic term and two quartic vertices, it gives rise to a non-dynamical matrix theory, where the fields are the four infinite-dimensional matrices associated to the $\mathcal A_\mu$. However the equation of motion \eqref{eq:motion} has to be further studied, because it can probably produce more interesting vacua, as the one found in \cite{Grosse:2007jy}. The vacuum problem in noncommutative field theory is indeed an interesting question. Notice that it has been studied for the scalar theory with harmonic term in \cite{deGoursac:2007uv}.

In the line of the non-vanishing vacuum question, some others problems remain to be understood and properly controlled as the gauge fixing and the ghost sector \cite{Grosse:2007qh}. After that, subsection 2.4 gives some lights about the choice of a convenient regularisation scheme for the study of renormalisability.

\vskip 1 true cm 

\noindent
{\bf{Acknowledgments}}: I would like to thank the organisers of the ``International Conference on Noncommutative Geometry and Physics'' for their invitation. I am also grateful to J.-C. Wallet and R. Wulkenhaar for their collaboration. \par

\end{document}